# K. Alex Müller and his important role in ferroelectricity

Annette Bussmann-Holder[1] and Hugo Keller[2]

[1] Max-Planck-Institute for Solid State Research, Heisenbergstr. 1, D-70569 Stuttgart, Germany

[2] Physik-Institut der Universität Zürich, Winterthurerstr. 190, CH-8057 Zürich, Switzerland

Abstract: In this review we concentrate on the work of K. Alex Müller in connection with his activities on oxide perovskites and ferroelectrics which were central to his research career long before he successfully discovered the first high temperature superconductor (HTSC) together with J. G. Bednorz in 1986. Not accidentally, but taking his long experience in perovskite ferroelectrics into account, the first HTSC was an oxide perovskite which had never been considered before to be superconducting.

My (ABH) first encounter with K. Alex Müller (Alex) was in 1979 in Portoroz, former Yugoslavia, on the occasion of the Fourth European Meeting on Ferroelectricity. At that time Alex was already a superstar in the ferroelectrics community and well-known for his sharp mind and quick comprehension. Compared to all following meetings the ferroelectrics community was of a manageable size and could be considered almost as a family - everyone knew everyone and conflicts and controversies between the different participants were well-known. As a result, fervid discussion rounds occurred after speeches, and for those who were not involved in those heated disputes, an amusing scenario was presented. K. Alex Müller was often involved in such discussions and redoubtable for his punctured responses. Actually, I was extremely impressed by him and looked up to him like to a diva.

The first time I (HK) heard the name of "K. Alex Müller" was 1971 when I was a student of physics at ETH Zurich. I was attending a lecture of Prof. W. Känzig on "Phase Transitions and Critical Phenomena", where the pioneering work of Alex and Walter Berlinger on the experimental determination of the static critical exponents $\beta$ of the order parameter of the second-order structural phase transitions in $SrTiO_3$ (STO) and $LaAlO_3$ was presented (see below). Then, I first met Alex at the University of Zurich, when, as a PhD student I attended a lecture of Alex on "Group Theory". Since my PhD thesis at the University of Zurich on the critical behavior of quasi two-dimensional magnetic systems with perovskite structure was related to a similar topic Alex was working on, he invited me to give a seminar at the IBM Zurich Research Laboratory. I was very proud to present my work in such a famous laboratory, and Alex at that time was already a world famous expert in the field of phase transitions and critical phenomena. A true scientific collaboration with Alex started shortly after his discovery of high-temperature superconductivity in the cuprates in 1987. Subsequently, I had the great privilege to collaborate with him on different aspects of cuprates, including: isotope effects, symmetry of the order parameter, intrinsic inhomogeneity, for over a period of almost 30 years.

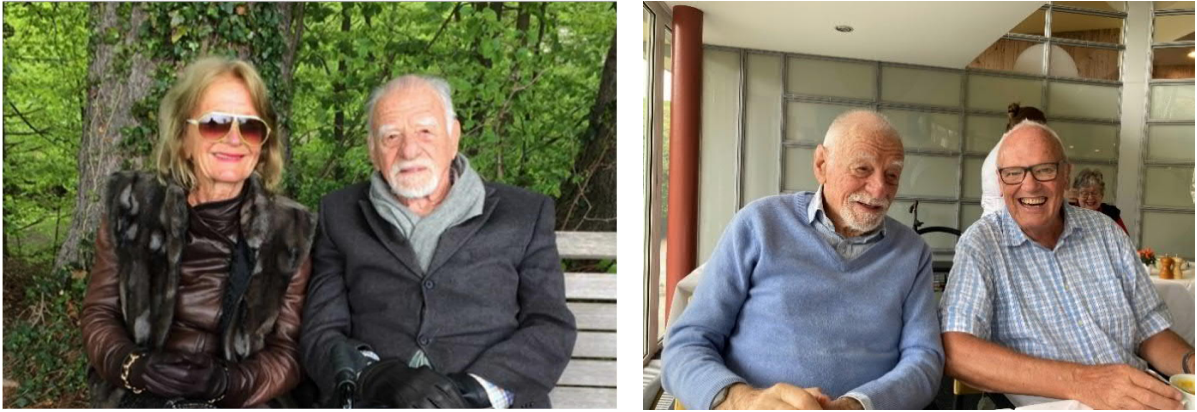

Figure 1 (left) Annette and Alex in front of the Tertianum (Zurich, 2019); (right) Alex and Hugo in the cafeteria of the Tertianum (Zurich, 2022).

I would like to start by elucidating how he became involved in the physics of perovskites and his career paths in the field. Alex studied physics at ETH Zürich where he was intensely influenced and impressed by the lectures of Prof. Wolfgang Pauli. His diploma work was on the Hall effect in grey tin which was supervised by Prof. G. Busch, who was also the supervisor of his PhD thesis in 1958 which dealt with the paramagnetic resonance in the newly synthesized double perovskite $SrTiO_3$ (STO) [1].

After finalizing his PhD thesis, Alex started his professional life as head of the magnetic resonance group at the Battelle Memorial Institute in Geneva. Upon the recommendation of Prof. E. Brun he did his habilitation at the University of Zurich in 1962. In view of his high impact in the scientific research in perovskites, the IBM Zurich Research Laboratory in Rüschlikon offered him in 1963 the position of a researcher where he was subsequently promoted to a group leader of the physics department in 1971, a position which he held until 1985. His research focused on STO and related perovskites with emphasis on their chemical binding, their ferroelectric and soft-mode properties, and later on their critical and multicritical phenomena at their phase transitions. In 1970 he was appointed as titular professor of the University of Zurich and a decisive moment in his career occurred in 1982 when he was nominated as IBM fellow. This enabled him to decide freely and independently about his further research areas – a milestone on his way to the Nobel prize.

The favorite experimental tool of Alex was Electron Paramagnetic Resonance (EPR), which had been discovered by the end of the Second World War. This methodology is rather cheap in comparison with related probes and allows testing of the local environment, crystalline surroundings and structural properties of specific ions. His first paper (his PhD thesis) was published in 1958 [1] entitled "Paramagnetic resonance of $Fe^{3+}$ in single crystals of $SrTiO_3$". In this work he demonstrated unambiguously and for the first time that a cubic to tetragonal structural phase transition occurs in $SrTiO_3$. He used EPR in a $Fe^{3+}$ doped STO single crystal with a doping concentration of $10^{17}/cm^3$. The spectra were taken at 3.2 cm wavelength at room temperature and 80K. At room temperature where STO is cubic, the splitting |3*a*| at zero magnetic field is $(5.95\pm0.30)10^{-4} cm^{-1}$. The magnitude of |3*a*| admits the conclusion that $Fe^{3+}$ occupies the $Ti^{4+}$ lattice position. At 80K, i.e. in the tetragonal phase, the lines $\pm5/2 \leftrightarrows \pm3/2$ and $\pm3/2 \leftrightarrows \pm1/2$ split if the magnetic field is along the [100] direction whereas this splitting is absent if the field is directed along [111]. This demonstrates that the STO single crystal consists of tetragonal domains below the phase transition temperature at $T_S=105K$. Note, that this finding is different for the one observed for the ferroelectric phase in other perovskites (see below).

These results together with EPR studies of manganese and $Cr^{3+}$ ions in STO, which also confirm the tetragonal phase of the STO, were published in [2, 3]. From there on he focused on this zone boundary related instability in oxide perovskites and showed that these are driven by the tetragonal rotation of $TiO_6$ octahedra in STO and the trigonal rotation of $AlO_6$ octahedra in $LaAlO_3$ below their respective phase transitions. The normalized rotation angles φ vary quantitatively in the same way as a function of reduced temperature. Thus φ is the order parameter in such structural phase transitions, which are characteristic for perovskite compounds and is defined as the rotation angle φ of the oxygen octahedra around the tetragonal c-axis in STO ($LaAlO_3$) below the transition temperature $T_s$ [4].

Related to this topic is a pioneering work of Alex and Walter Berlinger [5] on the experimental determination of the static critical exponent β of the order parameter φ of the second-order structural phase transition in STO and $LaAlO_3$. Just below $T_s$ the temperature dependence of φ follows the power law: $\varphi(t) \sim (1-t)^\beta$, where $t=T/T_s$ is the reduced temperature. In this study it is clearly shown that for temperatures very close to $T_s$ (t > 0.9) the order parameter is well described by a critical exponent β=0.33(2), demonstrating that the critical behavior of the order parameter φ is *universal*. However, outside the critical temperature region (0.7 < t < 0.9) the order parameter follows the classical mean-field Landau behavior with β =0.5. This impressive work is a real landmark in the field of phase transitions and critical phenomena.

Early on Alex was also interested in the Jahn-Teller (JT) effect which had been predicted by Jahn and Teller in 1937. Notably, using EPR its existence could be proven for paramagnetic impurities in crystals with orbital degenerate ground states. By investigating $Ni^{3+}$, $Cu^{2+}$, $Pt^{3+}$ impurities in STO and other oxides, the static as well as the dynamic JT effect were substantiated [6].

Aside from the zone boundary related instability in perovskites he was also working on the instability related to the zone center which is accompanied with a soft transverse optic mode. This was intensely investigated, especially in STO, since inelastic neutron scattering (INS) results clearly showed its existence and thereby the possibility of a polar instability towards ferroelectricity. Together with H. Burkard he measured the related dielectric constant down to 4K, much lower than had been done with INS and observed that the dielectric constant achieved huge values, however, started saturating towards a high constant value below ~30K. This finding led to his most cited paper (except those related to high temperature superconductivity) "SrTiO$_3$: An intrinsic quantum paraelectric below 4K" [7] where the authors proved that quantum fluctuations grow in size with decreasing temperature and become larger than the soft mode displacement. Since the citation to this work keeps on growing with time (Figure 2), even nowadays, it was named a "sleeping beauty".

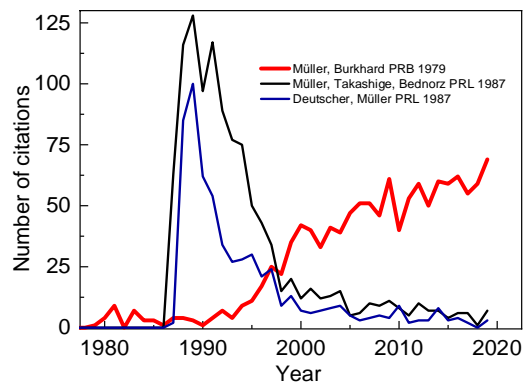

Figure 2 Number of citations with time of the most important papers of K. Alex Müller excluding the most cited one [8]. The red line shows the citations to Ref. 7. (Taken and modified from Ref. 9)

Together with J. Georg Bednorz he continued his research on this topic by looking into possible ferroelectric order in Ca doped STO [10]. Here Ca occupies the Sr position. Since its radius is much smaller than the one of Sr, ferroelectricity appears in the AO plane in ABO$_3$ due to off-centering of the Ca ion. With increasing doping a transition from an XY (planar) quantum ferroelectric to randomness occurs which vanishes completely with further Ca content. These results were a hint to a method to achieve ferroelectricity in an otherwise quantum paraelectric.

Another interesting observation of Alex, which suggests a possible origin of ferroelectricity in oxide perovskites, was based on the EPR analysis of $Fe^{3+}$ in various systems for which he used a superposition model [11]. In the model compound it was found that $Fe^{3+}$ in ferroelectric BaTiO$_3$ does not have ionic coordinates identical to $Ti^{4+}$. However, it remains always in the center of the oxygen octahedron, i.e. it does not move out of the inversion symmetric position as the $Ti^{4+}$ does and thus does not take part in the phase transitions of the host lattice. This makes it a very useful tool to probe the bulk oxygen lattice alone. This behavior of $Fe^{3+}$ in BaTiO$_3$ is not a single example, but rather appears to be a general feature in ferroelectric oxide perovskites, since a related behavior was also observed by him in PbTiO$_3$ and KNbO$_3$. This might explain why $Fe^{3+}$ doping in these perovskites always suppresses $T_c$. In addition, it supports the viewpoint that the spontaneous polarization of these materials stems from the shift of the $Ti^{4+}$, $Nb^{5+}$ ions from the center of the oxygen octahedra. The small differences in the mass of Ti versus Fe or their effective charge are thus not responsible for the different behavior. However, the $d^0$ configuration is crucial apart from the anisotropic nonlinear polarizability of the oxygen ions (absent for $S^{2-}$, $F^-$, $C^{1-}$, etc.). If the latter p electron charge can move onto empty d cation orbital, the oxygen shell polarizability is enhanced and ferroelectricity occurs.

For a while, Alex considered himself to be a single ion man, but Prof. Harry Thomas introduced him to many particle physics. From EPR experiments he concluded that the dynamics of the phase transitions of perovskites change in the vicinity of the transition. The critical properties of systems are on one hand characterized by the system's dimensionality, meaning linear in one dimension, two dimensional in planes, or three and hypothetical four dimensions, on the other hand by its symmetry class of interactions. It turned out that uniaxial pressure modifies the symmetry class of the phase transition in STO and consequently its critical behavior. These experiments enabled him to classify the transformation as Ising type, bicritical or cubic. It was thus possible to identify a critical end point as a function of temperature and pressure. His measurements were the first ones evidencing a three dimensional Potts behavior [12] whereby the system approaches in a jump like manner the transformation. The measured order-parameter discontinuity at the transition depends on the trigonal order parameter with exponent $δ=0.62(10)$ (mean-field theory predicts $δ=1$). This agrees with renormalization-group predictions, and proves that the model has a first-order transition even in the quantum fluctuation dominated region.

$BaTiO_3$ attracted increased interest by him, since it exhibits three successive phase transitions, which have been classified as order/disorder in the US whereas W. Cochran [13] attributed perfect mode softening to them, i.e. a displacive character. This latter description was generally accepted and the order/disorder aspects discarded. In 1985/86 Alex very successfully demonstrated that the transition from orthorhombic to rhombohedral is of order/disorder character by replacing the four valued titanium in part by the isovalent manganese [14]. This work is even today – 30 years later – considered as pioneering. Collaborations with Prof. G. Völkel [15] of the University of Leipzig and in more recent work with him on the same samples as used in 1985 highlighted these aspects and contributed to a further understanding of the order/disorder aspect.

When he was only 53 he had attained an enormous scientific reputation which could have been the end of his career as a researcher. However, as is well known, he continued with science with much greater success than ever before. Interestingly, his work on perovskite oxides together with his intimate knowledge of the JT effect gave rise to the discovery of high-temperature superconductivity [8] as is shown in the Fig. 3 below where the clear connection between both is apparent.

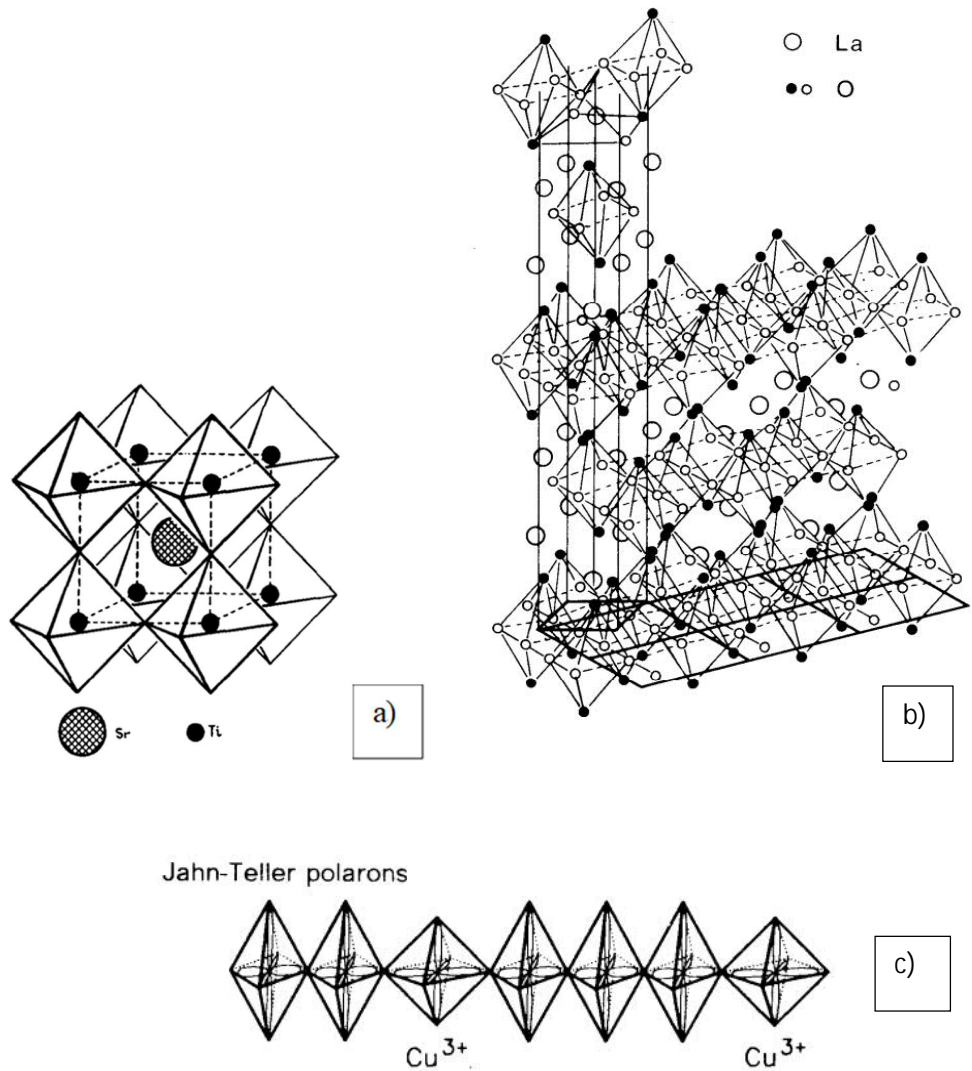

Figure 3. a) Crystal structure of STO. Figure taken from the PhD thesis of Alex [1]. b) Crystal structure of the cuprate superconductor $La_2CuO_4$. From [16]. c) The schematic concept of JT polarons. From [17]

Alex was a truly devoted teacher and kept contact with the Physik-Institut of the University of Zurich. This contact was not only on a scientific basis, but also for social means. He was not only interested in the research progress of the students but also in their lives and supported them financially through the *K. Alex Müller Foundation*.

Alex married his wife Inge in 1956, and a year later his son Erich was born, followed in 1960 by his daughter Sylvia. Three grandchildren stem from both. He was extremely devoted to his family and Inge always supported him, giving structure to his life. They enjoyed travelling together and she joined him on his numerous journeys.

Besides his family, Alex was very enthusiastic about old cars and especially Jaguars. In order to do repairs himself, his garage was equipped with a lower level from which he managed to work on the underside of the vehicles. In addition to his passion for exact natural sciences, he also showed broad interest in natural philosophy and in the depth psychology of C.G. Jung, as well as in classical music, literature, art, and history.

Another of Alex's lifelong hobbies was skiing which he had learned during his time in the Boarding School in Schiers (located in the Swiss alps). He became an excellent skier, constantly practiced and was mostly happy when he could stay in the mountains with his skis. Almost up to the age of 87 he continued with this sport and then only gave up because his muscles weakened with age. His children as well as theirs learned skiing from him until he confessed that they were doing better than him.

In addition to skiing, he also practiced swimming close to his house in Hedingen in a small lake regularly in the summer.

Alex has inspired the world and initiated the work on ferroelectricity. He was one of the greatest scientists we ever met, but in addition, he was a true friend of not only to us, but also our families as he took enormous interest in their development. We have lost a scientific father, a very good friend, an inspiring scientist, an everlasting teacher who was warm-hearted, friendly, open minded, sharp-witted, and had a large portion of black humor. Good bye, and take care!

His words to us: "Macht's gut, Euer Alex!" (which translates to, "Take it easy, yours Alex!")

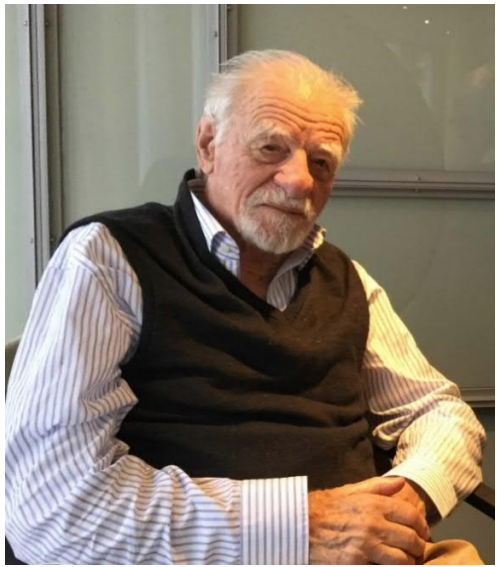

Figure 4 Alex in the Tertianum (Zurich, 2022)